\documentclass[prb,twocolumn,amsmath,amssymb,floatfix,superscriptaddress,aps]{revtex4-2}

\usepackage{color}
\usepackage{graphicx}
\usepackage{hyperref}
\usepackage{braket}
\usepackage[capitalize]{cleveref}
\usepackage[normalem]{ulem}
\usepackage{bbm}
\usepackage{soul}
\usepackage[normalem]{ulem}

\begin{document}
\title{Thermoelectric properties of interacting double quantum dots}
\author{Nahual Sobrino}\email{nsobrino@ictp.it}
\affiliation{The Abdus Salam International Center for Theoretical Physics (ICTP), Strada Costiera 11, 34151 Trieste, Italy}

\date{\today}

\begin{abstract}
	We investigate the thermoelectric transport properties of an interacting  parallel double quantum dot in the Coulomb-blockade regime. Building on an analytical solution based on an equation-of-motion technique, we extend the formalism for the asymmetrically coupled situation and provide compact closed-form expressions for steady-state currents together with the differential conductance, Seebeck coefficient, and thermal conductance. 
	  We determine the operating points that maximize efficiency and output power of the system,  clarifying their relation to standard near-equilibrium $ZT$ expressions. We further study the thermal rectification in both the open- and closed-circuit configurations and derive an expression for the open-circuit case. Interaction-induced resonances are understood in terms of the poles of the resulting Green's function,  generating gate and bias dependent regions of enhanced efficiency at finite power, negative differential thermal conductance, and finite thermal rectification.
\end{abstract}

 \maketitle
\section{Introduction}

The study of thermoelectric transport in nanoscale systems is of fundamental and practical interest because it allows for precise control over quantum confinement, Coulomb interactions, and nonequilibrium driving, with implications for energy conversion and thermal functionalities \cite{dresselhaus2007new, snyder2008complex,zebarjadi2012perspectives,bell2008cooling,shakouri2011recent,vineis2010nanostructured,perez2023thermoelectric, biele2019beyond, AGRAIT200381}. Among available systems, quantum dots (QDs) and, in particular, double quantum dots (DQDs), offer extensive tunability via gate voltages and contact couplings, enabling systematic access to charge states, correlation effects, and interference phenomena relevant to both charge and heat flow \cite{Zhang2023, alivisatos1996perspectives, kastner1993artificial, reed1988observation, bayer2001coupling, vanderwiel2002electron, hanson2007spins, petta2005coherent}.
This tunability makes DQDs excellent systems to investigate rectification and conversion mechanisms that lack  straightforward bulk analogues and serve as benchmarks for model-driven device design \cite{yadalam2019statistics,mazal2019nonmonotonic,matern2023metastability, vzitko2010fano,tanaka2005interference,zhang2021thermal,perez2025role, haug2008quantum,datta1995electronic,di2013nonequilibrium}.

On the theory side, multiple frameworks address the study of interacting nanostructures in transport setups in and out of equilibrium, including hierarchical equations of motion (HEOM), numerical renormalization group (NRG), quantum master equations (QME), and dynamical mean-field theory (DMFT) \cite{bhandari2021nonequilibrium, tanimura2020numerically,wilson1975renormalization,sobrino2021thermoelectric,li2005quantum,breuer2002theory,georges1996dynamical,sobrino2019steady,ryndyk2009green}.
Within Green’s function (GFs) approaches, the equation-of-motion (EOM) method provides a systematic route to deriving the equations governing the dynamics of the GFs  \cite{zubarev1960double,hubbard1963electron,hubbard1964electronII,hubbard1964electronIII}.
The EOM technique has been applied to single and multiple QD systems under various coupling and interaction regimes \cite{kang1995equation,van2010anderson,kang1998transport,swirkowicz2003nonequilibrium,sierra2016interactions,alomar2016coulomb,chang2008theory,sztenkiel2007electron,kuo2007tunneling,sun2002double,chi2006interdot},
\begin{figure}
	\centering
	\includegraphics[width=0.9\linewidth]{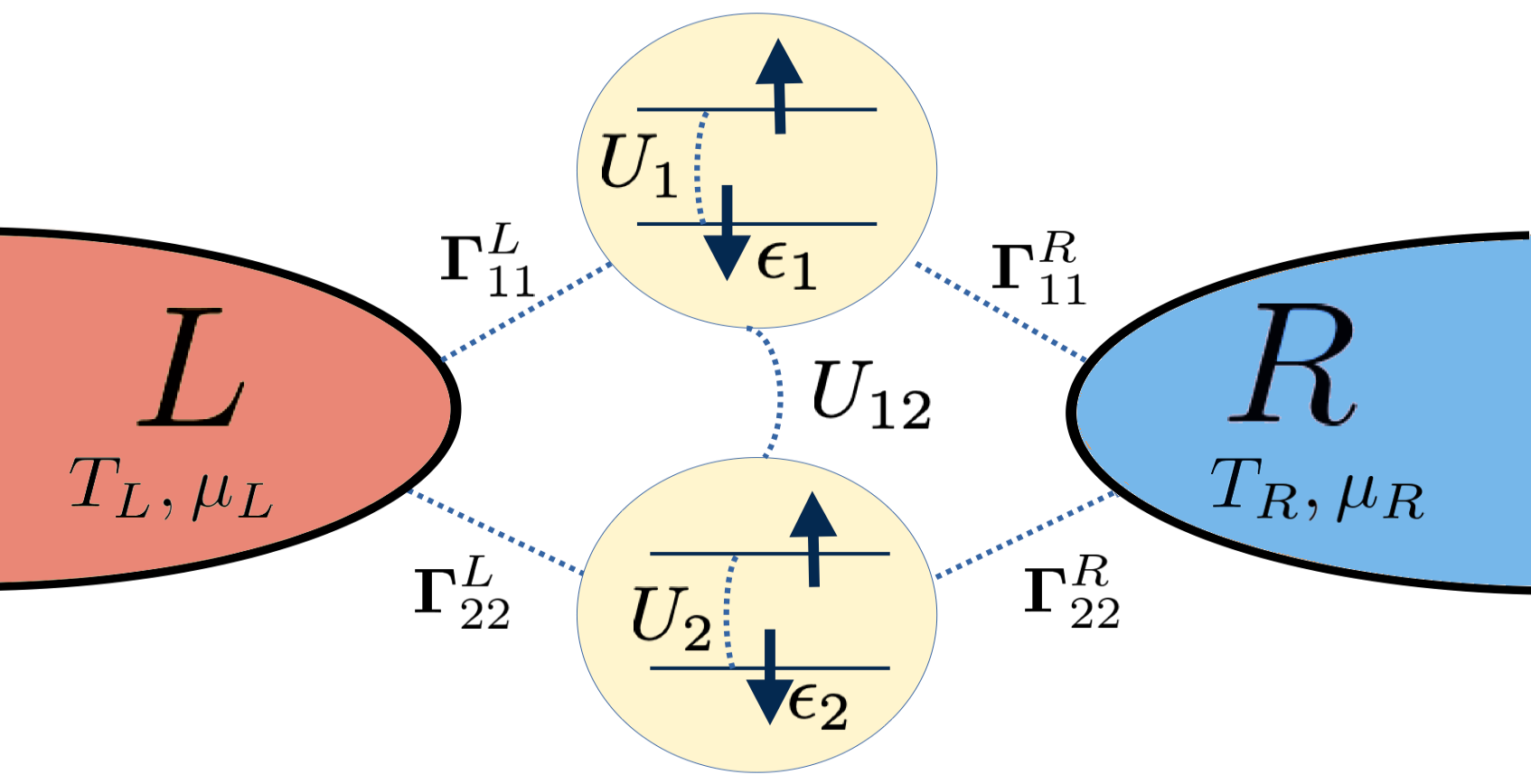}
	\caption{Schematic transport setup representation of the DQD system. The interacting dots are coupled to the leads at temperatures $T_\alpha$, and  chemical potentials $\mu_\alpha$ through the coupling strengths $\boldsymbol{\Gamma}_{ii}^{\alpha}=\gamma_\alpha/2$, with $\alpha=L,R$.}
	\label{fig1}
\end{figure}
but closed-form nonequilibrium expressions are scarce given the complex structure of the equations, and the solution is typically obtained through a self-consistent numerical procedure. Recently, an analytical derivation of the EOM for the DQD in
the Coulomb blockade regime in and out of equilibrium has been presented in Refs.~\cite{Sobrino2024,Sobrino2025}.  Complementing these theoretical developments, recent experiments on QDs have demonstrated controlled thermoelectric operation, including nonlinear thermocurrent, Kondo-enhanced thermopower, and energy harvesting functionality~\cite{thierschmann2015three,dorsch2021heat,josefsson2018quantum,svilans2018thermoelectric,svensson2013nonlinear}.

Despite these advances, some issues of direct practical relevance remain largely unexplored from an analytical perspective. The combined impact of Coulomb interactions and coupling asymmetry on charge and heat transport has not been established in closed fully-analytical form, and key performance metrics, like the thermal rectification,  and thermoelectric efficiency, as well as the differential transport coefficients, lack compact analytical expressions that make their dependence on gate voltage, bias, and other system parameters explicit. A concise and interpretable description of these dependencies is essential to identify, e.g. operating points of high efficiency at finite power  \cite{benenti2017fundamental,juergens2013thermoelectric,donsa2014double,zimbovskaya2020thermoelectric,sobrino2023thermoelectric,balduque2025quantum}, and an analytical description provides not only a controlled way of tracing the origin of the physical behavior, but a computationally efficient method to explore the parameter space.

In this work, we develop an analytical treatment of the thermoelectric properties of an interacting parallel DQD in the Coulomb-blockade regime beyond the linear response. Building on the previously derived EOM solution for the equilibrium case \cite{Sobrino2024},
we obtain closed-form expressions for steady-state charge and heat currents and for the \emph{differential} transport coefficients under finite temperature and voltage biases and asymmetric couplings. 
 We further analyze thermal rectification in closed and open circuit configurations, and evaluate thermoelectric efficiency (both its maximum value and its value at maximum power) highlighting parameter regimes where interaction-induced resonances and coupling asymmetry yield finite rectification and enhanced conversion at finite power.

The paper is organized as follows. Section~\ref{sec:model} introduces the model and summarizes the analytical EOM method for the steady state. Section~\ref{sec:currents_andtrasnp_coeffs} presents closed-form expressions for the currents and differential coefficients beyond linear response and discusses their relation to the pole structure of the Green’s function. Sections~\ref{sec:efficiency} and~\ref{sec:rectification} analyze thermoelectric efficiency and thermal rectification for representative parameters, emphasizing the roles of interactions and coupling asymmetry. Conclusions are given in \cref{sec:conclusions}. Finally, details of the EOM derivation are provided in \cref{app_Green_functions}.

\section{Model and Method}
\label{sec:model}
\subsection{Double Quantum Dot Hamiltonian}
The system of interest is schematically illustrated in \cref{fig1} and consists of a parallel double quantum dot attached to two electron reservoirs in local thermal equilibrium at temperatures $T_\alpha$ and chemical potentials $\mu_\alpha$, with $\alpha=L,R$. The Hamiltonian of the system is given by $\mathcal{H}=\mathcal{H}_0+\mathcal{H}_{leads}+\mathcal{H}_{tunn}$
where 
\begin{subequations}
\begin{align}
	\mathcal{H}_0 =& \sum_i\varepsilon_{i}\hat n_{i}+\sum_{i,\sigma} U_{i}\hat n_{i\sigma}\hat n_{i\bar\sigma}+U_{12}\hat n_{1}\hat n_{2}\;,\\
	\mathcal{H}_{leads}&=\sum_{\alpha i k \sigma}\epsilon_{\alpha  k i}\hat c^{\dagger}_{\alpha k i \sigma}\hat c_{\alpha  k i \sigma}\;,\\
	\mathcal{H}_{tunn} &= \sum_{i \alpha k \sigma }\left(V_{\alpha  k i}\hat c^{\dagger}_{\alpha  k i \sigma}\hat d_{i \sigma}+\text{H.c.}\right)\;,
\end{align}
	\label{eq_H}
\end{subequations}
being $\hat d^\dagger_{i\sigma}$ ($\hat d_{i\sigma}$) the  creation (annihilation) operator for an electron with spin $\sigma$ on dot $i$,  $\hat c^{\dagger}_{\alpha  k i \sigma}$ ($\hat c_{\alpha  k i \sigma}$)  the  creation (annihilation) operator for an electron with spin $\sigma$ in state $k$ of lead $\alpha$ coupled to dot $i$, and the total and spin-resolved density operators $\hat n_{i}=\hat n_{i\sigma}+\hat n_{i\bar\sigma}$ and $\hat n_{i\sigma}=\hat d^{\dagger}_{i\sigma}\hat d_{i\sigma}$. The isolated double dot is described by $\mathcal{H}_0$, with $\varepsilon_i$ and $U_{i}$ representing the on-site energy and the intra-Coulomb repulsion of dot $i$, respectively, and $U_{12}$ being the inter-Coulomb repulsion between the two dots. $	\mathcal{H}_{leads}$ describes the single-particle eigenstates of the leads and $\mathcal{H}_{tunn}$ corresponds to the tunneling between the dots and the two reservoirs, with $V_{\alpha  k i}$ the tunneling coupling parameter.  Throughout this work we consider diagonal tunneling couplings and we adopt the wide-band approximation is adopted: each reservoir is treated as a featureless continuum with chemical potential \(\mu_\alpha\), so that the corresponding embedding self-energy becomes energy-independent $
\Delta_{i}^{\alpha} \;=\; -\,\frac{i}{2}\,\gamma_{i}^{\alpha}$, 
where \(\gamma_{i}^{\alpha}\) is the level broadening induced by electrode \(\alpha\) on dot \(i\).  In addition, we focus on the regime of proportional coupling in which the linewidth matrices obey \(\boldsymbol{\Gamma}^{L} = \lambda\,\boldsymbol{\Gamma}^{R}\) with \(\lambda \in \mathbb{R}\) and 
\(\boldsymbol{\Gamma}^{\alpha}_{ij}  = \gamma_\alpha \delta_{ij}/2\).

A temperature difference \(\Delta T = T_L - T_R\) and/or a dc voltage bias \(V = \mu_L - \mu_R\) drives the junction away from equilibrium.  After transient effects have died out, the system reaches a steady state characterized by an electric current \(I\) and the associated energy and heat currents, \(W\) and \(Q\).  By convention, currents flowing into the central region are taken to be positive.  Charge and energy conservation then imply $I \equiv I_L = -\,I_R$, 
$W \equiv W_L = -\,W_R$,
$Q \equiv Q_L = P - Q_R$,
where $P$ is the electrical power $P = -\,I\,V$.


\subsection{Equation of Motion Approach}
\label{subsec:EOM}

The equations of motion (EOM) technique is used to obtain an analytic expression for the retarded one–body Green’s function (GF) of the interacting DQD,
\begin{equation}
	G^r_{i\sigma}(\omega)\equiv \langle\!\langle \hat d_{i\sigma} : \hat d^\dagger_{i\sigma}\rangle\!\rangle=\int_{-\infty}^{\infty} dt\,e^{i\omega t}\,G^r_{i\sigma}(t),
	\label{G_onebody}
\end{equation}
where $G^r_{i\sigma}(t)\equiv -\,i\,\theta(t)\,\big\langle\{\hat d_{i\sigma}(t),\hat d^\dagger_{i\sigma}(0)\}\big\rangle$. Here and throughout, the symbol ``$:$'' follows Zubarev’s notation \cite{zubarev1960double}: it separates the two operators inside \( \langle\!\langle \cdot \rangle\!\rangle \) and indicates the retarded structure defined in \cref{G_onebody}. For a generic operator $\hat B$, the EOM reads
\begin{equation}
	(\omega+i0^+)\,\langle\!\langle \hat B : \hat d^\dagger_{i\sigma}\rangle\!\rangle
	=\big\langle\{\hat B,\hat d^\dagger_{i\sigma}\}\big\rangle+\langle\!\langle [\hat B,\mathcal H] : \hat d^\dagger_{i\sigma}\rangle\!\rangle.
	\label{eq:EOM-generic}
\end{equation}
Successive application of \eqref{eq:EOM-generic} to operators $\hat B$ of the form $\hat B=\hat D\,\hat O$, with $\hat D=\prod_{\ell}\hat n_{i_\ell\sigma_\ell}$ a finite product of distinct local density operators and $\hat O\in\{\hat d_{i\sigma},\,\hat c_{\alpha k i\sigma}\}$, generates an infinite hierarchy that couples the single–particle GF to (i) dot–only higher–order GFs $\langle\!\langle \hat D\,\hat d_{i\sigma}:\hat d^\dagger_{i\sigma}\rangle\!\rangle$ and (ii) mixed dot–lead GFs such as $\langle\!\langle \hat D\,\hat c_{\alpha k i\sigma}:\hat d^\dagger_{i\sigma}\rangle\!\rangle$.

In the Coulomb–blockade (CB) regime, where the level broadening is small compared with $k_{\mathrm B}T$, the hierarchy can be closed at finite order while retaining \emph{all} dot–only GFs (no mean–field factorization of dot operators). This closure is consistent with the local–density commutation approximation $[\,\hat n_{i\sigma},\mathcal H\,]\!\approx\!0$, since $\hat n_{i\sigma}$ commutes with the isolated–dot and lead Hamiltonians and the tunneling contribution is linear in the hybridization and enters the EOM only beyond leading order. The approximation is exact as the hybridization vanishes and, at the level of one–body propagators, it amounts to broadening the isolated–dot poles while keeping the full dot–correlator structure. 

In the nonequilibrium steady-state, correlators appearing in \cref{eq:EOM-generic} are evaluated via
\begin{equation}
	\big\langle \hat D \hat n_{i'\sigma'}\big\rangle
	= -\int \frac{d\omega}{2\pi}\,\tilde f(\omega)\,
	\mathrm{Im}\,\langle\!\langle \hat D \hat d_{i'\sigma'} : \hat d^\dagger_{i'\sigma'}\rangle\!\rangle,
	\label{eq:NEQ-closure}
\end{equation}
where \(
f_\alpha(\omega)=[1+e^{(\omega-\mu_\alpha)/T_\alpha}]^{-1}
\)
and
$\tilde f(\omega)=\tfrac1\gamma\big[\gamma_L f_L(\omega)+\gamma_Rf_R(\omega)\big]$. Note that these correlators depend explicitly on both leads couplings, bias and temperatures through $\tilde f(\omega)$. \cref{eq:NEQ-closure} reduces to the exact equilibrium formula when $T_L=T_R$ and $\mu_L=\mu_R$.  The recursive evaluation of the GFs with \cref{eq:EOM-generic,eq:NEQ-closure} gives a \emph{finite} linear system for the independent correlators. Its analytic solution provides the local occupations $\langle \hat n_i\rangle$ and the higher–order correlators in closed form, see \cref{eq_solution_two_body_2}. Because all dot GFs can be written as  sums of simple poles, the frequency integrals collapse to combinations of 
\begin{align}
	\phi(p)
	&\equiv \int \frac{d\omega}{2\pi}\,\tilde f(\omega)\,\frac{\gamma}{(\omega-p)^2+\gamma^2/4} \nonumber\\
	&= \frac12-\sum_{\alpha=L,R}\frac{\gamma_\alpha}{\gamma\pi}\,\mathrm{Im}\,
	\psi\!\left(\frac12+\frac{\gamma/2+i(p-\mu_\alpha)}{2\pi T_\alpha}\right),
	\label{eq:phi}
\end{align}
with $\psi$ the digamma function and $\gamma=\gamma_L+\gamma_R$. The matrix elements of the correlator system are linear combinations of $\phi$ evaluated at the addition/removal energies. The single-particle retarded GF then takes the form
\begin{equation}
	G^r_{i\sigma}(\omega)=\sum_{j=1}^{6}\frac{r_{i,j}}{\omega-p_{i,j}+i\gamma/2},
	\label{eq:Gr-sixpole}
\end{equation}
with poles
\small
\begin{align*}
	p_{i,1}&=\varepsilon_i, &
	p_{i,2}&=\varepsilon_i+U_i, &
	p_{i,3}&=\varepsilon_i+U_i+U_{12}, \nonumber\\
	p_{i,4}&=\varepsilon_i+U_i+2U_{12}, &
	p_{i,5}&=\varepsilon_i+U_{12}, &
	p_{i,6}&=\varepsilon_i+2U_{12},
	\label{eq:poles}
\end{align*}
\normalsize
corresponding to the DQD addition/removal energies. The residues $r_{i,j}$ follow analytically from the same linear system and depend \emph{only} on the model parameters and external driving forces, see \cref{eq_numerators_one_body}. Further details of the analytical expression for the higher order GFs and the correlators can be found in \cref{app_Green_functions}.

It is worth mention that the derivation presented holds for the situation of diagonal tunneling rates and is restricted to the case in which the one body part of $\mathcal{H}_0$ involves the density operator of each dot separately. This includes for example a T-shaped DQD, as long as the tunneling term $t_{12}=0$. Inclusion of terms of the form $\hat c^\dagger_i \hat c_j$ or off-diagonal tunneling rates would lead to finite contributions of the off-diagonal GFs and off-diagonal correlators, increasing the EOM hierarchy. 
\section{Currents and Transport Coefficients}
\label{sec:currents_andtrasnp_coeffs}

\subsection{Charge and Heat Currents}

The steady–state charge and heat currents follow from the nonequilibrium Keldysh expressions \cite{meir1992}

\begin{align*}
	I =&   \frac{i}{4\pi}\int d\omega\,
	\Big\{
	\mathrm{Tr}\!\big[\big(f_L(\omega)\boldsymbol{\Gamma}^L-f_R(\omega)\boldsymbol{\Gamma}^R\big)\big(\mathbf{G}^r(\omega)-\mathbf{G}^a(\omega)\big)\big]
	\nonumber\\
	&+\mathrm{Tr}\!\big[\big(\boldsymbol{\Gamma}^L-\boldsymbol{\Gamma}^R\big)\mathbf{G}^{<}(\omega)\big]
	\Big\}, 
	\\
	Q =& \frac{i}{4\pi}\int d\omega\,(\omega-\mu_L)\,
	\Big\{
	\mathrm{Tr}\!\big[\big(f_L(\omega)\boldsymbol{\Gamma}^L-f_R(\omega)\boldsymbol{\Gamma}^R\big)\times\nonumber\\
	&\big(\mathbf{G}^r(\omega)-\mathbf{G}^a(\omega)\big)\big]
	+\mathrm{Tr}\!\big[\big(\boldsymbol{\Gamma}^L-\boldsymbol{\Gamma}^R\big)\mathbf{G}^{<}(\omega)\big]
	\Big\}
\end{align*}
In the situation of proportional coupling $\boldsymbol{\Gamma}^L=\lambda\,\boldsymbol{\Gamma}^R$, the currents simplify to 
\begin{subequations}
\begin{align}
	I &= \frac{1}{2\pi}\sum_i \int d\omega\,[f_L(\omega)-f_R(\omega)]\,\mathcal{T}_i(\omega), 
	\\
	Q &= -\frac{1}{2\pi}\sum_i \int d\omega\,(\omega-\mu_L)\,[f_L(\omega)-f_R(\omega)]\,\mathcal{T}_i(\omega),
\end{align}
\label{eq_currents_landauer}
\end{subequations}
with $\mathcal{T}_i(\omega)=-\frac{\tilde\gamma}{2}\,\mathrm{Im}\,G^r_{i\sigma}(\omega)$ and $\tilde\gamma=4\gamma_L\gamma_R/(\gamma_L+\gamma_R)$. Substituting the GF  \cref{eq:Gr-sixpole} yields fully analytic expressions 
\begin{subequations}
\begin{align}
	I &=
	-\frac{\tilde\gamma}{2\pi}\sum_{i,j,\alpha}s_\alpha\, r_{i,j}\;
	\mathrm{Im}\!\left[\psi\!\left(z_{i,j}^{\alpha}\right)\right],
\\[2pt]
	Q &=
	\frac{\tilde\gamma}{2\pi}\sum_{i,j\alpha} s_\alpha\, r_{i,j}
	\left[
	\frac{\gamma}{2}\,\mathrm{Re}\!\left[\psi\!\left(z_{i,j}^{\alpha}\right)\right]\right.\nonumber\\
	&\left.-\Big(p_{i,j}-\frac{V}{2}\Big)
	\mathrm{Im}\!\left[\psi\!\left(z_{i,j}^{\alpha}\right)\right]
	\right]+\frac{2\gamma_L\gamma_R}{\pi}\ln\!\left(\frac{T_L}{T_R}\right),
\end{align}
\label{eq_currents_digamma}
\end{subequations}
where $z_{i,j}^{\alpha}=\frac12+\frac{\gamma/2+i(p_{i,j}-\mu_\alpha)}{2\pi T_\alpha}$, $s_L=+1$, and $s_R=-1$. 
In \cref{eq_currents_digamma} the currents depend  explicitly on the local couplings, not only through $\tilde \gamma$ but also through $r_{ij}$ as given in \cref{eq_numerators_one_body}.

In the following, the analytical EOM results are used to derive transport coefficients, thermoelectric performance, and thermal rectification across representative interaction regimes. A Python implementation of the method is available in Ref.~\cite{github_DQD}. 
\subsection{ Linear Response Transport Coefficients }
\label{sec:currents_coeffs}
In the linear–response regime the stationary charge and heat currents follow
\begin{align}
	\begin{pmatrix} I \\[2pt] Q \end{pmatrix}
	=
	\begin{pmatrix}
		L_{11} & L_{12} \\[2pt]
		L_{21} & L_{22}
	\end{pmatrix}
	\begin{pmatrix} V/T \\[2pt] \Delta T/T^{2} \end{pmatrix},
	\label{eq_LR_IQ}
\end{align}
and, in the absence of magnetic fields and time–reversal breaking, $L_{12}=L_{21}$ by Onsager reciprocity~\cite{onsager1931reciprocal}.

The kinetic coefficients are obtained by differentiating the closed current formulas in \cref{eq_currents_digamma} with respect to $V$ and $\Delta T$ and then taking the equilibrium limit
\begin{subequations}
	\label{eq:Lij}
	\begin{align}
		L_{11}
		&= T\,\frac{\partial I}{\partial V}\big|_{\substack{V=0\\ \Delta T=0}}
		= \frac{\tilde\gamma}{4\pi^{2}}
		\sum_{i,j} r_{i,j}\;
		\mathrm{Im}\big[\,i\,\psi'(z_{i,j})\big],
		\\[4pt]
		L_{12}
		&= T^{2}\,\frac{\partial I}{\partial \Delta T}\big|_{\substack{V=0\\ \Delta T=0}}
		= \frac{\tilde\gamma}{4\pi^{2}}
		\sum_{i,j} r_{i,j}\;
		\mathrm{Im}\big[\,l^{0}_{i,j}\,\psi'(z_{i,j})\big],
		\\[4pt]
		L_{22}
		&= T^{2}\,\frac{\partial Q}{\partial \Delta T}\big|_{\substack{V=0\\ \Delta T=0}}
		= \frac{\tilde\gamma}{4\pi^{2}}
		\sum_{i,j} r_{i,j}\!
		\left[
		p_{i,j}\,\mathrm{Im}\big[l^{0}_{i,j}\,\psi'(z_{i,j})\big]
		\right.\nonumber\\
		&\left.-\frac{\gamma}{2}\,\mathrm{Re}\big[l^{0}_{i,j}\,\psi'(z_{i,j})\big]
		\right]
		+\frac{2\gamma_L\gamma_R\,T}{\pi}.
	\end{align}
\end{subequations}
where $	l^{0}_{i,j}=\frac{\gamma}{2}+i\,p_{i,j}$, $z_{i,j}=\frac{1}{2}+\frac{\tfrac{\gamma}{2}+i\,p_{i,j}}{2\pi T}$, and $\psi'(z)=\frac{d^{2}}{dz^{2}}\log\Gamma(z)$ denotes the trigamma function \cite{abramowitz1968handbook}. 
It is convenient to recast \cref{eq_LR_IQ} in terms of standard transport coefficients. At linear order, the electrical conductance $G$, the Seebeck coefficient $S$ (thermopower), and the electronic contribution to the thermal conductance $\kappa$ read
\begin{subequations}
	\label{eq:GSkappa}
	\begin{align}
		G &= \frac{\partial I}{\partial V}\big|_{\substack{V=0\\ \Delta T=0}} = \frac{L_{11}}{T},
		\\[2pt]
		S &= -\,\frac{\partial V}{\partial \Delta T}\big|_{\substack{I=0\\ Q=0}} = \frac{L_{12}}{T\,L_{11}},
		\\[2pt]
		\kappa &= \frac{\partial Q}{\partial \Delta T}\big|_{\substack{I=0\\ QT=0}}
		= \frac{1}{T^{2}}\!\left(L_{22}-\frac{L_{12}^{2}}{L_{11}}\right),
	\end{align}
\begin{figure*}
	\centering
	\includegraphics[width=0.8\linewidth]{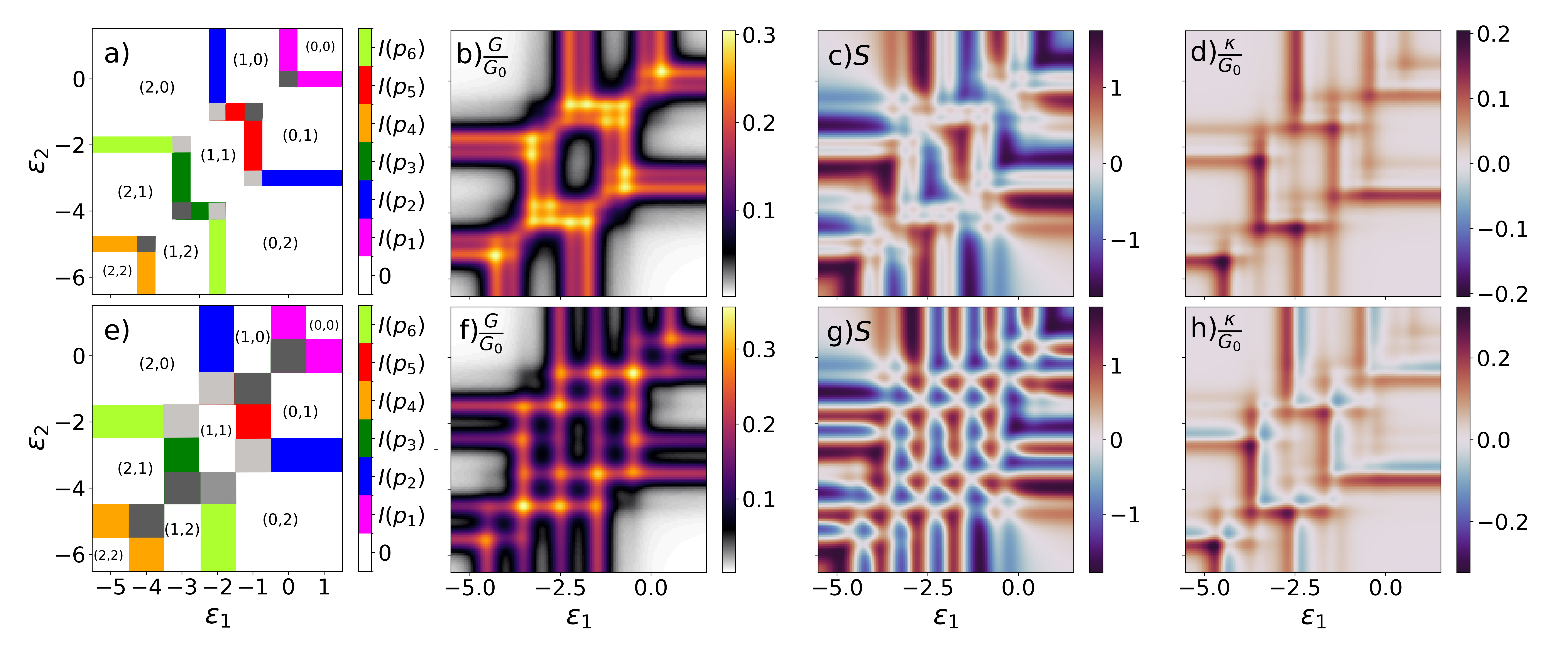}
	\caption{Differential  conductance $\frac{G}{G_0}$ (panels b), and f)), differential Seebeck $S$  (panels c), and g)), and differential thermal conductance $\frac{\kappa}{G_0}$  (panels d), and h)) as a function of the gates $\varepsilon_1$ and $\varepsilon_2$ for  $U_1=2$, $U_2=3$, $T=0.1$,  $\gamma = 0.1$, $\Delta T=0$, and up) $V=0.5$ down) $V=1$. Panels a) and e) show the finite charge current contributions from the poles ($p_j \in \{p_{1,j},p_{2,j}\}$) as defined in  \cref{eq_I_pi} in the low temperature limit. Energies in units of $U_{12}$.}
	\label{fig2}
\end{figure*}
\end{subequations}
From 	\cref{eq:GSkappa} one can directly obtain the standard figure of merit for thermoelectric performance
\begin{align}
ZT=\frac{S^{2}GT}{\kappa}
\end{align}
It increases with the power factor $S^{2}G$ and decreases with the thermal conductance $\kappa$, and it sets the linear–response upper bound for conversion efficiency (larger $ZT$ implies higher attainable efficiency).
Since the phonon heat transport (which would add $\kappa_{\mathrm{ph}}$) is neglected, the reported $ZT$ values represent upper bounds for a given parameter set.
\subsection{Transport Coefficients Beyond Linear Response}
Outside the linear regime,  where voltage and/or temperature differences are not small, linear‐response coefficients are not applicable; instead, thermoelectric properties are defined via \emph{differential} response functions at the chosen operating point $(V,\Delta T)$ \cite{dorda2016thermoelectric,manaparambil2023nonequilibrium}.
 Accordingly, the differential conductance is $G=\partial I/\partial V$, the differential Seebeck coefficient is $S=-\,\partial V/\partial\Delta T$, and the differential thermal conductance is $\kappa=\partial Q/\partial\Delta T$ (all derivatives taken at the same operating point).
These thermoelectric quantities are derived from  \cref{eq_currents_digamma} and take the form
\begin{subequations}
	\begin{align}
		G=&\frac{\tilde\gamma}{8\pi^2}\sum_{i,j,\alpha}\frac{ r_{i,j} }{T_\alpha}\text{Im}[i\psi'(z^{\alpha}_{i,j})] \; ,\\
		S=&\frac{-\tilde\gamma}{8\pi^2 G}\sum_{i,j,\alpha}r_{i,j }\text{Im}[l^{\alpha}_{i,j}\psi'(z^{\alpha}_{i,j})]\; ,
		\end{align}
			\begin{align}
		\kappa=&\frac{\tilde\gamma}{8\pi^2}\sum_{i,j,\alpha}r_{i,j}\left[ (p_{i,j}-V_L)\text{Im}[l^{\alpha}_{i,j}\psi'(z^{\alpha}_{i,j})] \right.\nonumber\\
		&\left.-\frac{\gamma}{2}\text{Re}[l^{\alpha}_{i,j}\psi'(z^{\alpha}_{i,j})]  \right]+\frac{8\gamma_{L}\gamma_R T}{\pi(4T^2-\Delta T^2)}\; ,
	\end{align}
	\label{eq_th_coef_BLR}
\end{subequations}
where $l^{\alpha}_{i,j}=\frac{\frac{\gamma}{2}+i(p_{i,j}-V_\alpha)}{T_\alpha^{2}}$%
, with $T_\alpha$ and $V_\alpha$ the temperature andbias voltage of lead $\alpha$.
 Beyond linear response the mixed derivatives need not coincide, i.e., $\partial I/\partial\Delta T \neq \partial Q/\partial V$. The explicit expression for the latter reads
\begin{align}
	\frac{\partial Q}{\partial V}=&\frac{\tilde\gamma}{8\pi^2}\sum_{i,j,\alpha}\frac{r_{i,j}}{T_{\alpha}}\left[ (p_{i,j}-V_L)\text{Im}[i\psi'(z^{\alpha}_{i,j})] \right.\nonumber\\
	&\left.-\frac{\gamma}{2}\text{Re}[i\psi'(z^{\alpha}_{i,j})]  \right]-\frac{I}{2}\;.
	\label{eq_dQdV_BLR}
\end{align}
Definitions beyond linear response  for the nonlinear Seebeck are not unique, and several operational choices have been explored in the literature \cite{azema2014conditions,eckern2020two,krawiec2007thermoelectric}. A common alternative is the \emph{open–circuit} Seebeck coefficient, obtained by imposing $I(V,\Delta T)=0$ and setting $S=V/\Delta T$ at that operating point.

In \cref{fig2}, the differential transport coefficients given by \cref{eq_th_coef_BLR} are shown under varying gates $\varepsilon_1$ and $\varepsilon_2$, and for different applied bias voltages between the electrodes $V=2V_L=-2V_R$. %
When a finite symmetric bias voltage is applied between the electrodes, the vertical and horizontal lines that define the regions of the stability diagram (and residues)  in the $(\varepsilon_1, \varepsilon_2)$ plane evolve due to the local bias dependencies in the Fermi functions $\tilde{f}$ of \cref{eq:phi}. If one makes the change of variable  $\omega'=\omega-V_\alpha$, the corresponding poles  will be shifted in an amount $V_\alpha$. In particular the poles of the GF related to the site $i$ will have an effective gate level $\varepsilon_i'=\varepsilon_i-V_\alpha$.
Since both the charge and heat current \cref{eq_currents_landauer} depend on the difference between the Fermi functions, the resulting contributions of each pole
cancel each other 
except in the stripe regions.  In the left panels of \cref{fig2} we
show in different colors the contribution of each of the poles  due to the different shift by the left and right bias in the low temperature limit.
The width of these regions exactly corresponds to the value of the applied voltage $V=V_L-V_R$, centered along the values of degenerate states at $V=0$. In particular, the $I(p_j)$ contributions correspond to the finite current generated due to the contribution of $p_j$
\begin{align}
	I(p_j)=& -\frac{\tilde\gamma}{2\pi} \sum_{i,\alpha}s_\alpha r_{i,j}\text{Im}[\psi(z^{\alpha}_{i,j})].
	\label{eq_I_pi}
\end{align}
With this definition, the total current is $I=\sum_j I(p_j)$. In particular, the poles related to site $i$ will generate the stripe regions along the direction where $\varepsilon_i$ varies. In \cref{fig2} the differential transport coefficients are shown for $V=0.5$ (upper panels) and $V=1$ (lower panels). In the low temperature limit, the  charge current
is essentially constant along the stripe regions, and therefore the differential conductance and Seebeck will be finite in the borders of the stripe regions. On the other hand, the heat current varies along the perpendicular direction of the stripes, and the thermal conductance will be finite in those regions. At higher temperatures, the differential transport coefficients are broadened due to temperature effect. Interestingly, one can observe that as the bias is increased, a negative contribution to the differential thermal conductance (NDTC) appears. This NDTC has a finite contribution inside the stripe regions, close to the side where the gate related to the other site is decreased.
\begin{figure*}
	\centering
	\includegraphics[width=1\linewidth]{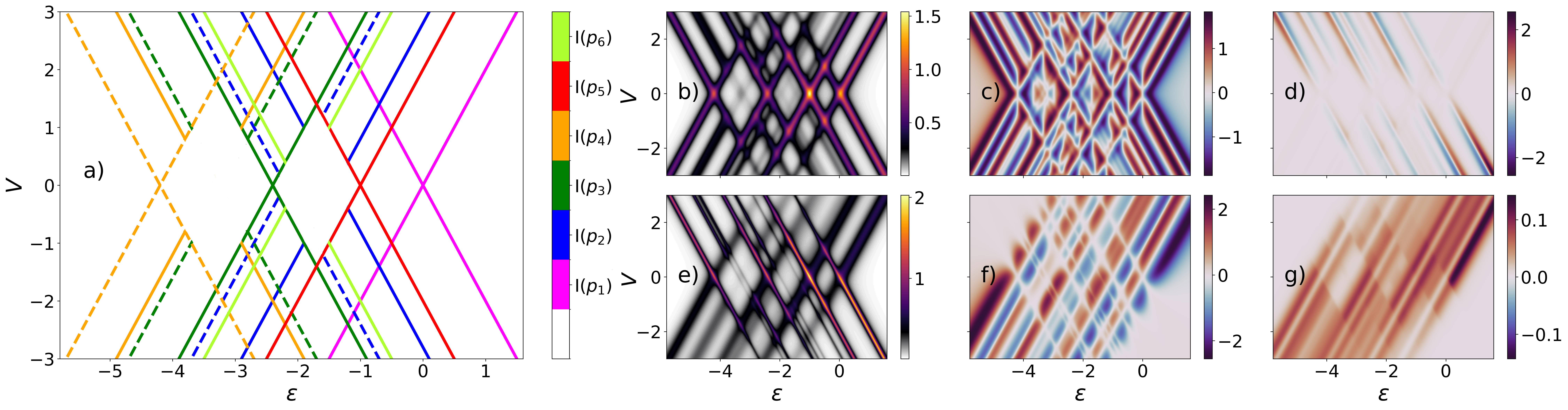}
	\caption{Differential  conductance $\frac{G}{ G_0}$ (panels b), and e)), differential Seebeck $S$  (panels c), and f)), and differential thermal conductance $\frac{\kappa}{ G_0}$  (panels d), and g)) as a function of the gate $\varepsilon$ and the bias $V$ for  $U_1=2.2$, $U_2=1.4$, $T=0.1$,  $\gamma = 0.1$, and up) $\Delta T=0$ down) $\Delta T\to 2T$. Panel a) shows the finite contributions from the poles ($p_j \in \{p_{1,j},p_{2,j}\}$)  corresponding to solutions of $p_{i,j} -V_\alpha = 0$, where the solid (dashed) lines correspond to site two (one) and the lines with positive (negative) slopes correspond to $V_L$ ($V_R$). Energies in units of $U_{12}$.}
	\label{fig3}
\end{figure*}
In \cref{fig3} the differential transport coefficients from \cref{eq_th_coef_BLR} are presented as a function of the common gate level $\varepsilon=\varepsilon_1=\varepsilon_2$ and the bias voltage $V=2V_L=-2V_R$, for $U_1=2.2$, $U_2=1.4$ in the case of equal temperatures and for large thermal gradients. The local temperatures are defined as $T_L=T+\Delta T/2$ and $T_R=T-\Delta T/2$, being $T$ the average temperature. In panel a) of \cref{fig3} a schematic representation of the positions of the poles contributions with finite residues from the single-particle Green's function is depicted,  where the solid (dashed) lines correspond to site two (one). The line crossings given by $p_{i,j} -V_\alpha = 0$ in the gate-bias plane, conform the structure of the charge and heat currents (see \cite{Sobrino2025}). This structure is responsible for the shape of the differential transport coefficients. In the upper panels of  \cref{fig3}  the differential transport coefficients are presented for $\Delta T=0$. The differential conductance is finite on the lines defined by $p_{i,j} -V_\alpha = 0$, reaching its maximum in the intersection between the lines. The differential Seebeck becomes finite in the sides of the poles lines where the residues are constant, reaching its maximum (minimum) in the left (right) side. Finally, the differential thermal conductance only shows a significant finite contribution due to the right lead contribution ($p_{i,j} -V_R = 0$), since in the symmetric applied bias configuration the left contribution is canceled due to the explicit $V_L$ term appearing as a prefactor of $\text{Im}[l^{\alpha}_{i,j}\psi'(z^{\alpha}_{i,j})]$ in \cref{eq_th_coef_BLR} c). In the lower figures of \cref{fig3} a finite thermal gradient $\Delta T\to 2T$ ($T_R\to 0$) is considered. In this limit the right–lead contribution to the differential conductance remains finite and becomes the main contribution, described by  a Lorentzian centered around the bias of the low temperature reservoir
\begin{equation}
	G_R\big|_{T_R\!\to0}=\frac{\tilde\gamma}{4\pi}\sum_{i,j} r_{ij}\,
	\frac{\tfrac{\gamma}{2}}{\big(\tfrac{\gamma}{2}\big)^2+\big(p_{ij}-V_R\big)^2}.
\end{equation}
For the Seebeck coefficient the large–argument expansion of $\psi'$ shows that the leading term of $l^R_{ij}\psi'(z^R_{ij})$ is purely real and therefore drops out after taking the imaginary part and the dominant visible weight in $S$ comes from the left–lead term evaluated at $T_L=2T$.
 Similarly, for the thermal response $\kappa$ the potentially divergent right–lead piece from the second line of \cref{eq_th_coef_BLR} c) cancels, leaving only a finite left–lead contribution
\begin{align}
	\kappa\big|_{T_R\!\to0}&=
	\frac{\tilde\gamma}{8\pi^2}\sum_{i,j} r_{ij}\!\left[
	(p_{ij}-V_L)\,\text{Im}\!\big[l^L_{ij}\psi'(z^L_{ij})\big]\right.\nonumber\\
	&\left.-\frac{\gamma}{2}\,\text{Re}\!\big[l^L_{ij}\psi'(z^L_{ij})\big]\right]_{T_L=2T}
	+\frac{\gamma_L\gamma_R}{\pi T_L},
\end{align}
with no net contribution from the cold right lead.
\begin{figure}
	\centering
	\includegraphics[width=0.8\linewidth]{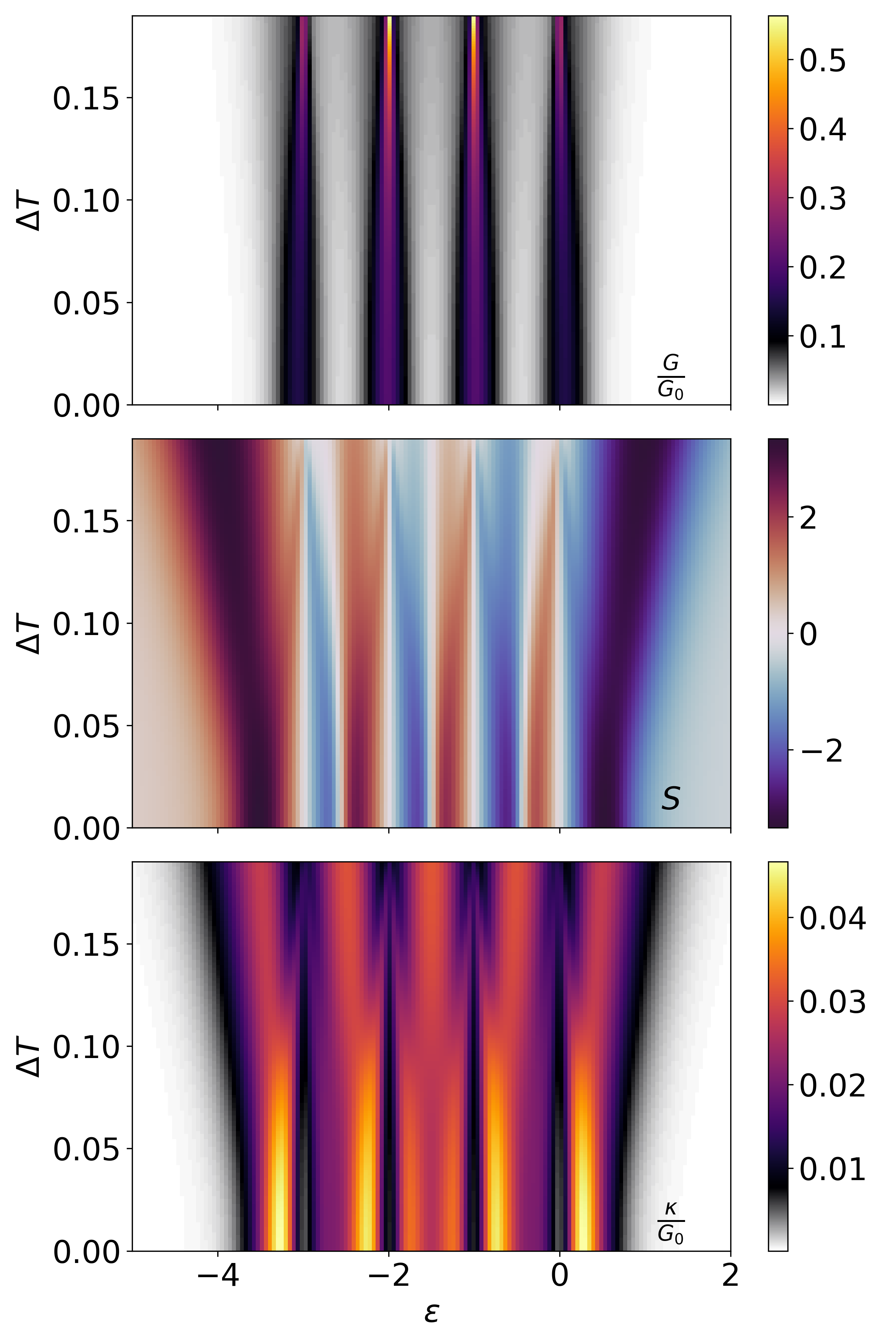}
	\caption{Differential transport coefficients as a function of the gate level $\varepsilon$ and the thermal gradient $\Delta T$ for  $U_1=U_2=1$, $V=0$, $T=0.1$, and $\gamma = 0.05$. Energies in units of $U_{12}$.}
	\label{fig4}
\end{figure}

In \cref{fig4}  the differential transport coefficients are presented as a function of the common gate level $\varepsilon=\varepsilon_1=\varepsilon_2$ and the thermal gradient $\Delta T$, for the constant interaction model (CIM) ($U_i=U_{ij}=1$) and $V=0$. The main features are structured along four main lines governed by $p_{1}=p_{3}=p_{4}=p_{5}=0$ (in the CIM case $p_{1,j}=p_{2,j}$ as long as $\varepsilon_1=\varepsilon_2$). This can be understood from \cref{fig3} a) if one takes a linecut at $V=0$. The four main peaks of the  differential conductance become steeper as the thermal gradient is increased, while the differential Seebeck alternates positive and negative values along the sides of the lines dictated by the poles, spreading more in the gates as the thermal gradient is increased. The differential thermal conductance shows small contributions on the lines of the poles, reaching its maximum value for relatively small thermal gradients on the sides of the lines.

\section{Thermoelectric efficiency}
\label{sec:efficiency}
The DQD system can act as a heat engine, converting thermal energy into electrical work. The efficiency of such a thermoelectric device is defined as
\begin{align}
	\eta = \frac{P}{Q_+}
	\label{eq_eta}
\end{align}
where $P = -IV$ is the electrical output power and $Q_+$ is the heat input from the hot reservoir. Two interesting situations to study are the maximum thermoelectric efficiency $\eta_{max}$ and the efficiency at maximum output power $\eta(P_{max})$.

The interest in $\eta_{max}$ stems from its role in understanding the theoretical limits of thermoelectric performance. Achieving $\eta_{max}$ implies the system is operating at its most efficient point, converting the maximum possible fraction of absorbed heat into useful work. 
In the linear response regime, where the currents follow the expressions \cref{eq_LR_IQ}, the system with fixed $\Delta T$ will achieve the maximum efficiency at the bias $V_{0}^{\text{LR}} =S\Delta T(1+ZT-\sqrt{1+ZT})/ZT$ obtained from $d\eta/dV=0$. The maximum thermoelectric efficiency around the linear response can be easily expressed in terms of the figure of merit by replacing the linear response expression for the currents \cref{eq_LR_IQ} evaluated at $V_0$ into \cref{eq_eta}
\begin{align}
	\eta_{max}^{\text{LR}} = \eta_C \frac{\sqrt{ZT + 1} - 1}{\sqrt{ZT + 1} + 1}
	\label{eta_max_LR}
\end{align}

where $\eta_C = 1- \frac{\Delta T}{T}$ is the Carnot efficiency. For our system, $\eta \to \eta_C$ only when $\gamma \to 0$ and $P \to 0$, indicating that the system will generate electricity at an infinitely slow rate, reflecting operation in a reversible cycle.

To proceed beyond linear response, we derive the thermoelectric efficiency using the current expressions obtained from the EOM approach \cref{eq_currents_digamma}. The bias at which the efficiency will be maximum $V_0$ is obtained from the equation ($Q_+=Q$ for $\Delta T>0$)
\begin{align}
	\left.\frac{d\eta}{dV}\right|_{V=V_0}=\frac{V_0}{Q}\left(\frac{I}{Q}\frac{\partial Q}{\partial V}-G\right)-\frac{I}{Q}=0
	\label{eq_V0}
\end{align}
where $G$, $S$, and $\frac{\partial Q}{\partial V}$ are the differential transport coefficients \cref{eq_th_coef_BLR,,eq_dQdV_BLR} to be evaluated at $V_0$ (as $I$ and $Q$). The maximum efficiency is therefore obtained by \cref{eq_eta} at the currents evaluated at $V_0$ resulting from solving self-consistently \cref{eq_V0}.

On the other hand, the efficiency at maximum output power, $\eta(P_{max})$, is equally important as it provides insights into the practical performance of thermoelectric devices under real-world operating conditions. This efficiency represents a balance between high output power and reasonable efficiency, reflecting the optimal operational state for applications where both power and efficiency are critical. This involves finding the bias voltage $V_1$ that maximizes the power $P$, and then evaluating the efficiency $\eta$ at this operating point. Mathematically, this can be approached by first determining $V_1$ through
\begin{align}
	\left. \frac{dP}{dV}\right|_{V=V_1} =-I-GV_1= 0,
	\label{eq_V1}
\end{align}
and subsequently calculating the corresponding efficiency $\eta(P_{max})$. In the linear response regime, the analytical expression for $\eta(P_{max})$ is given by:
\begin{align}
	\eta^{\text{LR}}(P_{\max}) = \frac{\eta_C}{2} \frac{ZT}{ZT + 2}.
	\label{eta_Pmax_LR}
\end{align}

In practical applications, maximizing $\eta$ and $\eta(P_{max})$ requires careful tuning of the system parameters such as the gate voltages, the Coulomb interactions $U_i$, and the tunneling couplings $\gamma$. The advantage of our approach lies in the ability to compute $\eta_{max}$ and $\eta(P_{max})$ beyond the linear response regime using the analytical solutions of the currents and differential transport coefficients and only $V_0$ and $V_1$ from  numerical evaluation. This allows us to have full control of the parameters and their dependence in modelling the thermoelectric performance under realistic conditions, providing valuable insights for optimizing the design and operation of DQD-based thermoelectric devices. 

\begin{figure}
	\centering
	\includegraphics[width=1\linewidth]{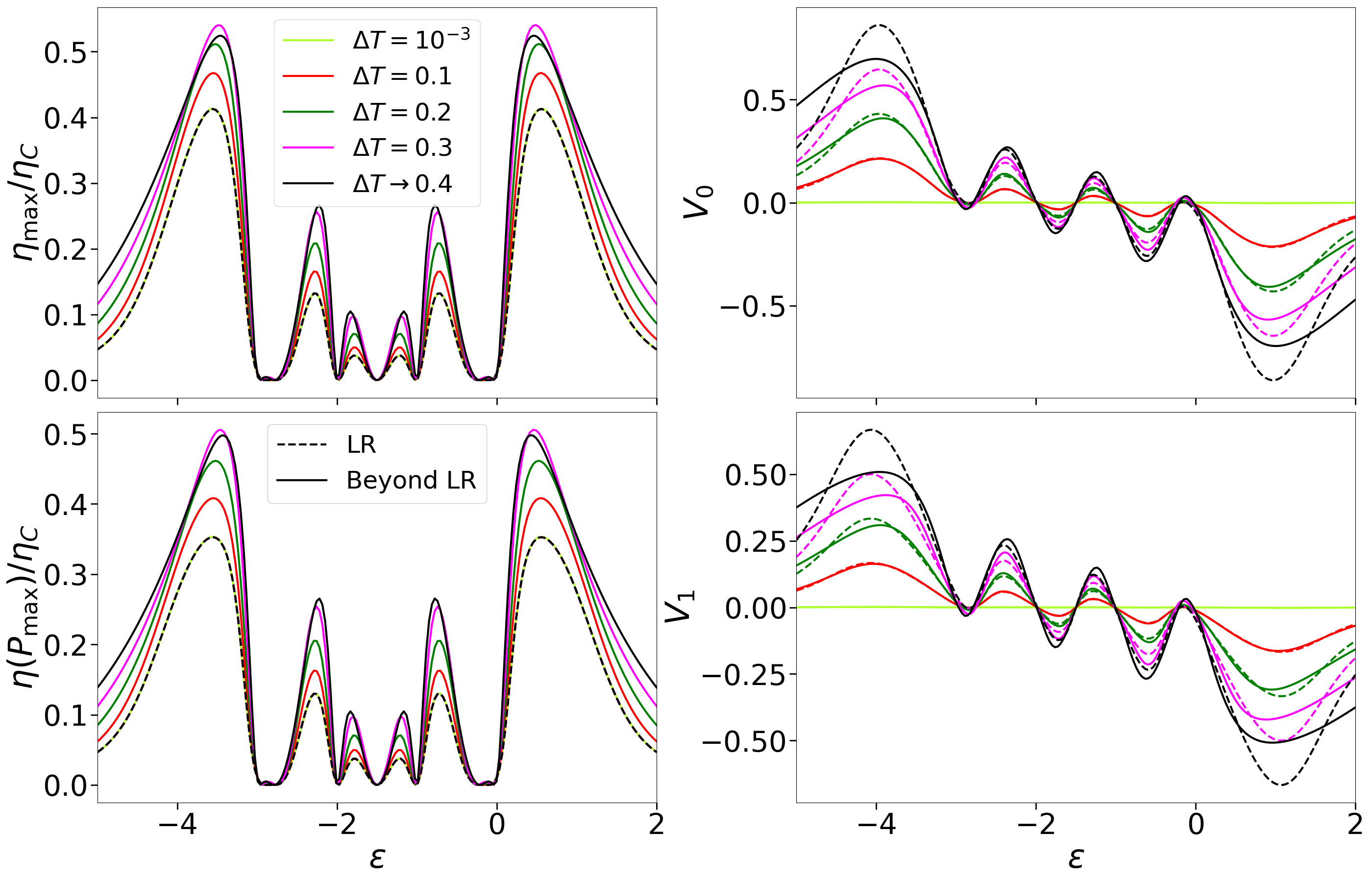}
	\caption{Up: maximum thermoelectric efficiency $\eta_{max}$ and bias $V_0$ that maximizes the efficiency, and down: efficiency at maximum output power $\eta(P_{max})$ and bias $V_1$ that maximizes the output power as a function of the gate level $\varepsilon=\varepsilon_1=\varepsilon_2$ for different fixed thermal gradients. The dashed lines represent the results obtained from linear response, while the solid line corresponds to the expressions beyond the linear response. The parameters are $T=2\gamma=0.2$, and $U_i=1$. Energies in units of $U_{12}$. }
	\label{fig5}
\end{figure}
The maximum thermoelectric efficiency and thermoelectric efficiency at maximum output power are presented in \cref{fig5} as a function of the gate level for different values of the thermal gradient. The dashed lines, corresponding to the linear response expressions  \cref{eta_max_LR,eta_Pmax_LR}, coincide with those obtained from the evaluation of \cref{eq_eta} with the currents \cref{eq_currents_digamma} for small thermal gradients. As the thermal gradient is increased and the system evolves beyond the linear response, the efficiencies preserve the structure of alternated maxima and minima as the gate is varied, but their values increase considerably. The reduced maximum efficiency found in the intermediate regions can be understood from the linear response, where the efficiency follows \cref{eta_max_LR}, being a monotonically increasing function of ZT. Around the (1,1) density region the transport window is governed by different resonance lines corresponding to the poles of the GF, producing a change of sign and a partially cancelation of  $S$,  while the differential conductance $G$ and the thermal conductance $\kappa$ remain similar to the (0,0) and (2,2) charge regions.  In the upper (lower) right panels of \cref{fig5} the bias $V_0$ ($V_1$) that maximizes the efficiency (output power) obtained from \cref{eq_V0} (\cref{eq_V1}) is compared against its linear response expression $V_0^{\text{LR}} = -\frac{L_{12}}{L_{11}} \frac{\Delta T}{T} \frac{1 + ZT - \sqrt{1 + ZT}}{ZT}$  ($V_1^{\text{LR}} = - \frac{L_{12}}{2L_{11}} \frac{\Delta T}{T}$) \cite{benenti2017fundamental}.

\begin{figure}
	\centering
	\includegraphics[width=1\linewidth]{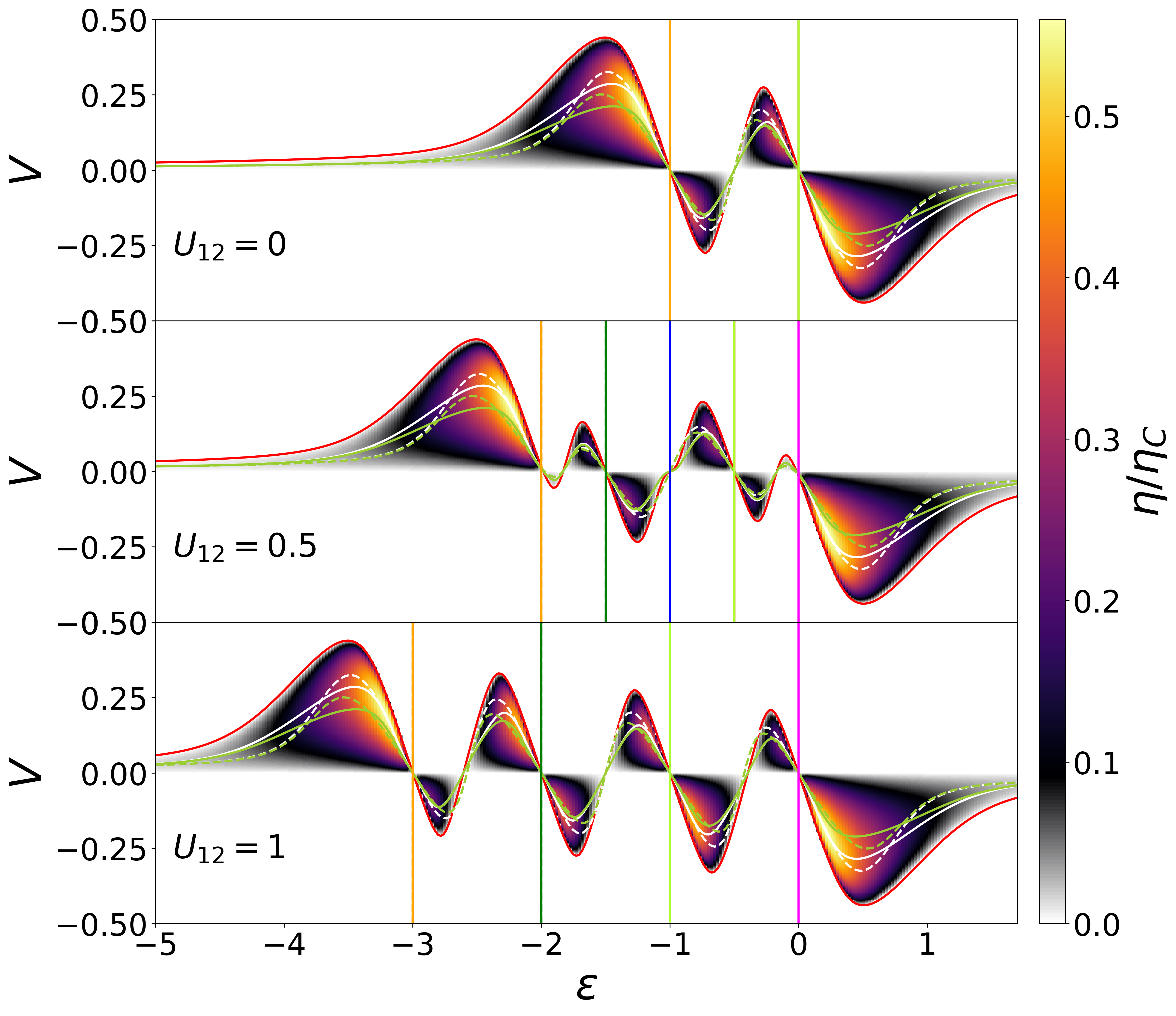}
	\caption{Thermoelectric efficiency $\eta/\eta_C$ as a function of the gate $\varepsilon$ and bias $V$ for three different values of $U_{12}$. The rest of the parameters are $U_1=1$, $T=2\gamma=0.1$, and $\Delta T=0.15$. Energies in units of $U_2$. The vertical lines correspond to the poles contributions with the same color code as in \cref{fig2}. The red line curve is the open circuit voltage $V_{oc}$. The green and the white solid  line curves correspond to $V_0$ and $V_1$ obtained by solving \cref{eq_V0} and \cref{eq_V1} respectively, while the dashed curves correspond to the linear response expressions $V_0^{\text{LR}} $ and $V_1^{\text{LR}} $.}
	\label{fig6}
\end{figure}
The thermoelectric efficiency can also be studied as a function of the bias for a given thermal gradient. In \cref{fig6} it is shown for three different values of the inter-dot Coulomb repulsion $U_{12}$ for  $U_1=U_2=1$, $T=2\gamma=0.1$, and $\Delta T=0.15$. The positive contributions to the thermoelectric efficiency are symetrically distributed around the particle-hole symmetric point in regions of positive and negative biases, inside the area defined by the open circuit voltage $V_{oc}$ (red line curve) and $V=0$. The gate at which these regions merge is given by  the pole contributions $p_{i,j}$ (corresponding to the vertical lines with the same color code as in \cref{fig3}). As the inter-dot Coulomb repulsion is increased, the pole contributions $p_{i,3},p_{i,4},p_{i,5},$ and $p_{i,6}$ are also increased, leading to a separation of the structures that define the regions of positive thermoelectric efficiency.  The green and the white solid  line curves correspond to $V_0$ and $V_1$ obtained by solving \cref{eq_V0} and \cref{eq_V1} respectively, while the dashed curves correspond to the linear response expressions $V_0^{\text{LR}} $ and $V_1^{\text{LR}} $.
\begin{figure*}
	\centering
	\includegraphics[width=0.8\linewidth]{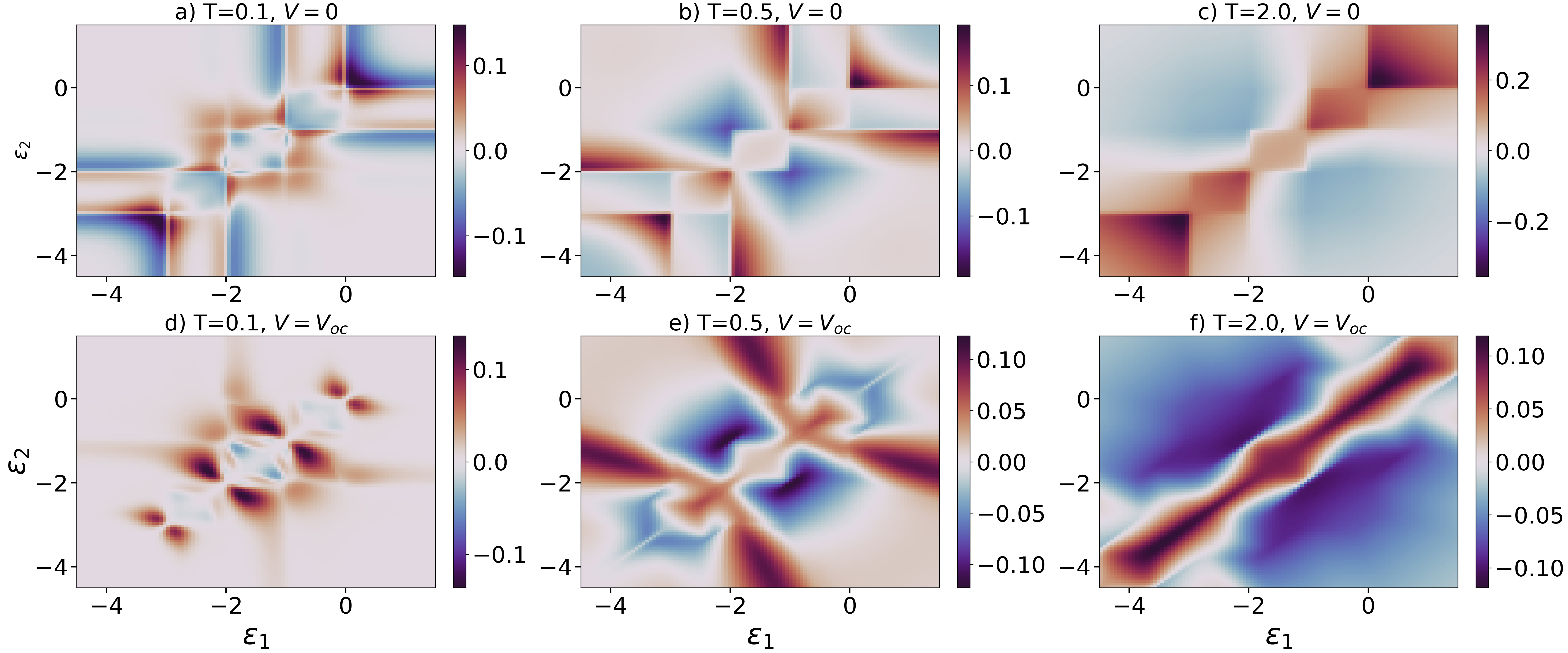}
	\caption{Up) Closed circuit ($V=0$) and down) Open circuit ($V=V_{oc}$) thermal rectification $R$ as a function of the gate levels $\varepsilon_1$ and $\varepsilon_2$ for different temperatures: left) $T=0.1$, center) $T=0.5$, right) $T=2$. The rest of the parameters are $\Delta T\to2T$, $\gamma_L=5\gamma_R=0.05$, and $U_1=U_2=1$. Energies in units of $U_{12}$.}
	\label{fig8}
\end{figure*}

\section{Thermal Rectification}
 \label{sec:rectification}
The thermal rectification is  a measure of the asymmetry in heat transport when the temperature gradient is reversed, and is quantified by the rectification coefficient $R$ \cite{tesser2022heat}, defined as
\begin{align}
	R=	\frac{|Q_+|-|Q_-|}{|Q_+|+|Q_-|}
\end{align}
where \(Q_+\) and \(Q_-\) are the heat currents for $T_L > T_R$ and $T_R > T_L$, respectively. Given the geometry of our system, a necessary condition for finite thermal rectification is an asymmetry in the couplings.
 In the closed-circuit configuration (\(V=0\)), where the forward and backward heat currents satisfy \(Q_{+}>0\) and \(Q_{-}<0\), 
the rectification coefficient can be written in compact form as
\begin{align}
	R 
	= \frac{\displaystyle \sum_{i,j} \xi_{i,j}\,\big(r_{i,j}^+ - r_{i,j}^-\big)}
	{\displaystyle \sum_{i,j} \xi_{i,j}\,\big(r_{i,j}^+ + r_{i,j}^-\big)
		+ 2\gamma \ln\!\left(\frac{T+\tfrac{\Delta T}{2}}{\,T-\tfrac{\Delta T}{2}\,}\right)} ,
	\label{eq:rectification_closed}
\end{align}
where $r_{i,j}^\pm$ are obtained from 
\cref{eq_numerators_one_body} evaluated at 
\((T_L,T_R)=(T\pm\tfrac{\Delta T}{2},\,T\mp\tfrac{\Delta T}{2})\), and the coefficients 
$	\xi_{i,j} = \sum_{\alpha} s_\alpha 
	\left[ \tfrac{\gamma}{2}\,\mathrm{Re}\,\psi\!\left(z_{i,j}^\alpha\right) 
	- p_{i,j}\,\mathrm{Im}\,\psi\!\left(z_{i,j}^\alpha\right) \right],
$
are evaluated at \((T_L,T_R)=(T+\tfrac{\Delta T}{2},\,T-\tfrac{\Delta T}{2})\).
In the asymmetric limit $\Delta T=2T$ (so that $T_\text{cold}\to 0$ and $T_\text{hot}=2T$), the expression reduces to
\begin{align}
	R\Big|_{\Delta T\to 2T}
	= \frac{\displaystyle \sum_{i,j} \Lambda_{i,j}\,\big(r_{i,j}^+ - r_{i,j}^-\big)}
	{\displaystyle \sum_{i,j} \Lambda_{i,j}\,\big(r_{i,j}^+ + r_{i,j}^-\big)} ,
\end{align}
with
\begin{align}
	\Lambda_{i,j} = 
	\tfrac{\gamma}{2}\!\left[\mathrm{Re}\,\psi\!\left(\tfrac{1}{2} + \tfrac{\gamma/2 + i p_{i,j}}{4\pi T}\right) 
	+ \ln\!\big(\frac{4\pi T}{\sqrt{(\frac{\gamma}{2})^2+p_{i,j}^2}}\big)\right] \nonumber\\
	- p_{i,j}\!\left[\mathrm{Im}\,\psi\!\left(\tfrac{1}{2} + \tfrac{\gamma/2 + i p_{i,j}}{4\pi T}\right) - \arctan (2p_{i,j}/\gamma)\right].
\end{align}
This form makes explicit that in the $T_\text{cold}\to 0$ limit the apparent logarithmic divergence  of \cref{eq:rectification_closed} is exactly canceled by the cold-lead expansion of $\psi$, so that $R$ remains finite.
 In the open-circuit situation, the bias $V$ is determined by solving for $V_{oc}=V(I=0)$ from \cref{eq_currents_digamma}. Here, despite the absence of a net charge current, the heat current remains finite due to the extra energy term $(\omega-V_L)$ in the integrand of \cref{eq_currents_landauer} b).

In \cref{fig8} the thermal rectification $R$ is plotted as a function of the gate levels $\varepsilon_1$ and $\varepsilon_2$ for different temperatures: left) $T=0.1$, center) $T=0.5$, right) $T=2$ and in the up) Closed circuit ($V=0$) and down) Open circuit ($V=V_{oc}$) setups for $\Delta T\to2T$, $\gamma_L=5\gamma_R=0.05$, and $U_1=U_2=1$. The finite contribution structure resembles the structure of the stability diagram for the CIM case \cite{sobrino2020exchange}, similar to the schematic representation illustrated in \cref{fig2}. In the closed circuit setup and at low temperatures (upper left panel) the thermal rectification presents positive and negative values in the regions of finite heat current. As the temperature is increased, $R$ spreads in the $(\varepsilon_1,\varepsilon_2)$ plane forming a structure where the maximum values are reached close to the lines of degenerate energy. This originates from the different behaviour of the heat currents $Q_\pm$ that define the rectification. As temperature increases, the transitions between occupation regions (set by the positions of the poles) associated with $Q_+$ are completely smoothed out, and the occupations evolve smoothly, acquiring non-integer values. By contrast, the occupations associated with $Q_-$ still display abrupt changes near the regions defined by the poles of the GFs. Therefore, this temperature-induced change in the occupations directly affects the residues (see \cref{eq_numerators_one_body}), and thus the rectification via the transmission function.
In the open circuit setup, the situation is more complicated, since at each $(\varepsilon_1,\varepsilon_2)$ point, the open-circuit bias $V_{oc}$ is different. In the low-temperature situation (lower left panel), the main rectification contributions are located along four regions close to the diagonal lines corresponding to degenerate states in the low bias situation. In both cases, when the temperature is large enough, the maximum thermal rectification is obtained close to the line $\varepsilon_1=\varepsilon_2.$

\section{Conclusions}
\label{sec:conclusions}
We studied the thermoelectric transport properties of asymmetrically coupled  double quantum dots. Using the equation-of-motion technique  we  provided compact closed-form expressions valid in the Coulomb blockade regime for the steady-state charge and heat currents together with the differential conductance, Seebeck coefficient, and thermal conductance, in terms of digamma and trigamma functions. The linear-response limit is recovered transparently, while the nonequilibrium situation remains described by the addition/removal energies (poles) of the one-body Green’s function. 

The gate- and bias-resolved transport response is governed by resonance lines \(p_{i,j}-V_\alpha=0\), corresponding to  the alignment of the excitation energies with the electrochemical potentials of the leads, which gives rise to the characteristic finite bias stripe structure. Within these stripes we identify regions of negative differential thermal conductance. In the strongly asymmetric thermal limit ($T_R\!\to\!0$) the right-lead contribution to the differential conductance becomes a simple Lorentzian centered at the cold-lead bias, while the potentially divergent pieces of the thermal response cancel, leaving a finite closed expression for $\kappa$. 

We studied the thermoelectric performance beyond linear response by determining the operating points that maximize (i) the efficiency and (ii) the output power via self-consistent conditions on the bias. Near equilibrium, the standard $ZT$-controlled expressions are recovered; at finite biases, the efficiency retains the pole-organized structure, exhibiting alternating maxima/minima across gate voltage with values that exceed their linear-response estimates.

Thermal rectification was analyzed in both closed circuit ($V\!=\!0$), where we derived a compact analytic expression for the rectification coefficient, and open circuit ($V\!=\!V_{\!oc}$), where a finite heat current persists despite $I\!=\!0$. At low temperatures the thermal rectification follows the stability diagram structure with pronounced positive and negative values localized along ground-state degeneracy lines at low temperature and, as temperature increases, broaden into neighboring regions of the $(\varepsilon_1,\varepsilon_2)$. For large temperatures, the maximum thermal rectification is obtained  in the region around $\varepsilon_1=\varepsilon_2.$
\appendix
\section{Green functions and  correlator expressions from EOM}
\label{app_Green_functions}

In this appendix we provide the closed form expressions of the GFs and the density correlators of the DQD in the asymmetric coupled and non-equilibrium case valid in the Coulomb-blockade regime. For a detailed equivalent derivation of  the symmetric equilibrium case see Ref.~\cite{Sobrino2024}. The formal difference in the derivation corresponds to  the evaluation of the density correlators: In the present situation, they are given by \cref{eq:NEQ-closure} and depend explicitly on both leads couplings, bias and temperatures through $\tilde f(\omega)$.
The recursive evaluation of the different order GFs with the anticommutation relations for the creation and annihilation operators	$\{\hat d_{i\sigma},\hat d_{j\sigma}^{\dagger}\}=\delta_{ij}$, $\{\hat d_{i\sigma},\hat d_{j\sigma}\}=\{\hat d_{i\sigma}^{\dagger},d_{j\sigma}^{\dagger}\}=0$ through the EOM \cref{eq:EOM-generic} allows to express them explicitly in the form of addition of single poles with the residues corresponding to linear combinations of the density correlators
\small
\begin{widetext}
	\begin{subequations}
	\begin{align}
		\braket{\braket{\hat n_{i\bar\sigma}\hat n_{\bar i \sigma }\hat d_{i\sigma}:\hat d^{\dagger}_{i\sigma}}}&=
		\frac{\braket{\hat n_{i\bar\sigma}\hat n_{\bar i\sigma}} - \braket{\hat n_{i\bar\sigma}\hat n_{\bar i\sigma}\hat n_{\bar i\bar \sigma}}}{\omega-\varepsilon_i - U_i-U_{12}+ i\frac{\gamma}{2}}
		+\frac{\braket{\hat n_{i\bar\sigma}\hat n_{\bar i\sigma}\hat n_{\bar i\bar \sigma}}}{\omega-\varepsilon_i-U_i-U_{12}+ i\frac{\gamma}{2}},\\
		\braket{\braket{\hat n_{\bar i\bar\sigma}\hat n_{\bar i \sigma }\hat d_{i\sigma}:\hat d^{\dagger}_{i\sigma}}}&=
		\frac{\braket{\hat n_{\bar i\bar\sigma}\hat n_{\bar i\sigma}}-\braket{\hat n_{i\bar\sigma}\hat n_{\bar i\sigma}\hat n_{\bar i\bar \sigma}}}{\omega-\varepsilon_i-U_i-U_{12}+ i\frac{\gamma}{2}}
		+\frac{\braket{\hat n_{i\bar\sigma}\hat n_{\bar i\sigma}\hat n_{\bar i\bar \sigma}}}{\omega-\varepsilon_i-U_i-U_{12}+ i\frac{\gamma}{2}},\\
		\braket{\braket{\hat n_{i\bar\sigma}\hat n_{\bar i \bar \sigma }\hat d_{i\sigma}:\hat d^{\dagger}_{i\sigma}}}&=
		\frac{\braket{\hat n_{i\bar\sigma}\hat n_{\bar i\bar \sigma}}-\braket{\hat n_{i\bar\sigma}\hat n_{\bar i\sigma}\hat n_{\bar i\bar \sigma}}}{\omega-\varepsilon_i-U_i-U_{12}+ i\frac{\gamma}{2}}
		+\frac{\braket{\hat n_{i\bar\sigma}\hat n_{\bar i\sigma}\hat n_{\bar i\bar \sigma}}}{\omega-\varepsilon_i-U_i-U_{12}+ i\frac{\gamma}{2}},\\
		\braket{\braket{\hat n_{i\bar\sigma}\hat d_{i\sigma}:\hat d^{\dagger}_{i\sigma}}}&
		= \frac{\braket{\hat n_{ i\bar\sigma}} - \braket{\hat n_{ i\bar\sigma}\hat n_{ \bar i\sigma}} - \braket{\hat n_{ i\bar \sigma}\hat n_{ \bar i\bar\sigma}} +
			\braket{\hat n_{i\bar\sigma}\hat n_{\bar i\sigma}\hat n_{\bar i\bar \sigma}}}{\omega-\varepsilon_i-U_i+ i\frac{\gamma}{2}}
		+ \frac{\braket{\hat n_{ i\bar \sigma}\hat n_{ \bar i \sigma}} +
			\braket{\hat n_{ i\bar \sigma}\hat n_{ \bar i \bar \sigma}}-
			2 \braket{\hat n_{i\bar\sigma}\hat n_{\bar i \sigma}\hat n_{\bar i\bar \sigma}}}{\omega-\varepsilon_i-U_i - U_{12}+ i\frac{\gamma}{2}}\nonumber\\&
		+ \frac{\braket{\hat n_{i\bar\sigma}\hat n_{\bar i\sigma}\hat n_{\bar i\bar \sigma}}}{\omega-\varepsilon_i-U_i - 2 U_{12} + i\frac{\gamma}{2}},\\
		\braket{\braket{\hat n_{\bar i \sigma}\hat d_{i\sigma}:\hat d^{\dagger}_{i\sigma}}}&
		= \frac{\braket{\hat n_{\bar i \sigma}}- \braket{\hat n_{ i\bar\sigma}\hat n_{ \bar i\sigma}} - \braket{\hat n_{ \bar i\sigma}\hat n_{ \bar i\bar\sigma}} +
			\braket{\hat n_{i\bar\sigma}\hat n_{\bar i\sigma}\hat n_{\bar i\bar \sigma}}}{\omega-\varepsilon_i-U_{12}+ i\frac{\gamma}{2}}
		+  \frac{\braket{\hat n_{i\bar\sigma}\hat n_{\bar i \sigma}} - \braket{\hat n_{i\bar\sigma}\hat n_{\bar i\sigma}\hat n_{\bar i\bar \sigma}}}{\omega-\varepsilon_i-U_i-U_{12}+ i\frac{\gamma}{2}}\nonumber\\
		&+  \frac{\braket{\hat n_{\bar i\sigma}\hat n_{\bar i \bar \sigma}} - \braket{\hat n_{i\bar\sigma}\hat n_{\bar i\sigma}\hat n_{\bar i\bar \sigma}}}{\omega-\varepsilon_i-2U_{12}+ i\frac{\gamma}{2}} 
		+  \frac{\braket{\hat n_{i\bar\sigma}\hat n_{\bar i\sigma}\hat n_{\bar i\bar \sigma}}}{\omega-\varepsilon_i-U_i-2U_{12}+ i\frac{\gamma}{2}},\\
		\braket{\braket{\hat n_{\bar i \bar \sigma}\hat d_{i\sigma}:\hat d^{\dagger}_{i\sigma}}}&
		= \frac{\braket{\hat n_{\bar i \bar \sigma}}- \braket{\hat n_{ i\bar\sigma}\hat n_{ \bar i\sigma}} - \braket{\hat n_{ \bar i\sigma}\hat n_{ \bar i\bar\sigma}} +
			\braket{\hat n_{i\bar\sigma}\hat n_{\bar i\sigma}\hat n_{\bar i\bar \sigma}}}{\omega-\varepsilon_i-U_{12}+ i\frac{\gamma}{2}}
			\nonumber\\
			&
		+  \frac{\braket{\hat n_{i\bar\sigma}\hat n_{\bar i \sigma}} - \braket{\hat n_{i\bar\sigma}\hat n_{\bar i\sigma}\hat n_{\bar i\bar \sigma}}}{\omega-\varepsilon_i-U_i-U_{12}+ i\frac{\gamma}{2}}
		+  \frac{\braket{\hat n_{\bar i\sigma}\hat n_{\bar i \bar \sigma}} - \braket{\hat n_{i\bar\sigma}\hat n_{\bar i\sigma}\hat n_{\bar i\bar \sigma}}}{\omega-\varepsilon_i-2U_{12}+ i\frac{\gamma}{2}} 
		+  \frac{\braket{\hat n_{i\bar\sigma}\hat n_{\bar i\sigma}\hat n_{\bar i\bar \sigma}}}{\omega-\varepsilon_i-U_i-2U_{12}+ i\frac{\gamma}{2}},
						\end{align}
		\begin{align}
		\braket{\braket{\hat d_{i\sigma}:\hat d^{\dagger}_{i\sigma}}}
		&= \frac{%
			\displaystyle 1-\braket{\hat n_{i\sigma}}
			-2\braket{\hat n_{\bar i\sigma}}
			+\braket{\hat n_{\bar i\sigma}\hat n_{\bar i\bar\sigma}}
			+2\braket{\hat n_{i\bar\sigma}\hat n_{\bar i\bar\sigma}}
			-\braket{\hat n_{i\bar\sigma}\hat n_{\bar i\bar\sigma}\hat n_{\bar i\sigma}}%
		}{\omega-\varepsilon_i + i\frac{\gamma}{2}}
      + \frac{%
			\displaystyle \braket{\hat n_{i\sigma}}
			-2\braket{\hat n_{i\bar\sigma}\hat n_{\bar i\bar\sigma}}
			+\braket{\hat n_{i\bar\sigma}\hat n_{\bar i\bar\sigma}\hat n_{\bar i\sigma}}%
		}{\omega-\varepsilon_i-U_i + i\frac{\gamma}{2}}
		 \nonumber\\
		 &+ \frac{%
			\displaystyle 2\!\left(\braket{\hat n_{i\bar\sigma}\hat n_{\bar i\bar\sigma}}
			-\braket{\hat n_{i\bar\sigma}\hat n_{\bar i\bar\sigma}\hat n_{\bar i\sigma}}\right)%
		}{\omega-\varepsilon_i-U_i-U_{12} + i\frac{\gamma}{2}}
		 + \frac{%
			\displaystyle \braket{\hat n_{i\bar\sigma}\hat n_{\bar i\bar\sigma}\hat n_{\bar i\sigma}}%
		}{\omega-\varepsilon_i-U_i-2U_{12} + i\frac{\gamma}{2}}
      + \frac{%
			\displaystyle \braket{\hat n_{\bar i\sigma}\hat n_{\bar i\bar\sigma}}
			-\braket{\hat n_{i\bar\sigma}\hat n_{\bar i\bar\sigma}\hat n_{\bar i\sigma}}%
		}{\omega-\varepsilon_i-2U_{12} + i\frac{\gamma}{2}}\nonumber\\
		&+ \frac{%
			\displaystyle 2\!\left(\braket{\hat n_{\bar i\sigma}}
			-\braket{\hat n_{\bar i\sigma}\hat n_{\bar i\bar\sigma}}
			-\braket{\hat n_{i\bar\sigma}\hat n_{\bar i\bar\sigma}}
			+\braket{\hat n_{i\bar\sigma}\hat n_{\bar i\bar\sigma}\hat n_{\bar i\sigma}}\right)%
		}{\omega-\varepsilon_i-U_{12} + i\frac{\gamma}{2}}.
	\end{align}
	\label{App_GFs_corrs}
	\end{subequations}
\end{widetext}
\normalsize
 The correlators appearing in the r.h.s of \cref{App_GFs_corrs} can be obtained from the GFs through
\cref{eq:NEQ-closure}, leading to a linear system with solution 
			\begin{subequations}
				\begin{align}
					&\braket{\hat n_{\bar i\sigma}\hat n_{\bar i\bar\sigma}}=\tau_{i,1}\braket{\hat n_{\bar i\sigma}},	
				\\
					&\braket{\hat n_{i\bar\sigma}\hat n_{\bar i\bar\sigma}}=\tau_{i,2}\braket{\hat n_{\bar i\sigma}},
\\
					&	\braket{\hat n_{i\bar\sigma}\hat n_{\bar i\bar\sigma}\hat n_{\bar i\sigma}}=\tau_{i,3}\braket{\hat n_{\bar i\sigma}},
					\\
					&	\braket{\hat n_{i\sigma}\hat n_{i\bar\sigma}\hat n_{\bar i\sigma}} =\tau_{i,4}\braket{\hat n_{\bar i\sigma}},
					\\
					& \braket{\hat n_{i\sigma}} =\frac{\phi(\varepsilon_i)\eta_{\bar i\bar i}-\phi(\varepsilon_{\bar i})\eta_{i\bar i}}{\eta_{11}\eta_{22}-\eta_{12}\eta_{21}},
				\end{align}
				\label{eq_solution_two_body_2}
			\end{subequations}
			where 
			\begin{subequations}
				\begin{align*}
					\eta_{ii}=&1+\phi(p_{i,1})-\phi(p_{i,2}),\\
					\eta_{i\bar i}=& \phi(p_{i,1})\left(2+\tau_{i,3}-\tau_{i,1}-2\tau_{i,2} \right) + \phi(p_{i,2})\left(2\tau_{i,2}-\tau_{i,4} \right)\nonumber\\
					+& 2\phi(p_{i,3}) \left(\tau_{i,3}-\tau_{i,2} \right)-\tau_{i,3} \phi(p_{i,4})\nonumber\\
					+& 2\phi(p_{i,5})\left(\tau_{i,1}+\tau_{i,2}-\tau_{i,3}-1 \right) + \phi(p_{i,6})\left(\tau_{i,3}-\tau_{i,1} \right),
				\end{align*}
				\label{eq_eta_params}
			\end{subequations}
			$\phi(p)$ given in \cref{eq:phi}, and
\begin{subequations}
	\begin{align*}
		\tau_{i,0} &= A_i\big[ D_i(1 - F_i) - F_i E_i \big]\nonumber\\
		&+ \big( 1 + \phi(p_{i,5}) - \phi(p_{i,3}) \big)\big[ 1 - F_i E_i C_{\bar{i}} \big], \\[4pt]
		\tau_{i,1} &= \frac{1}{\tau_{i,0}}\left(	\phi(p_{\bar{i},2})\big(\phi(p_{i,3}) - 1\big)\big(\phi(p_{i,5}) - 1\big)
		+ \phi(p_{i,5}\right) \, B_i) , \\[4pt]
		\tau_{i,2} &= \frac{1}{\tau_{i,0}}	\phi(p_{\bar{i},2})\big[ F_i(E_i + D_i) - D_i \big]
		+ \phi(p_{i,5}) \big[ 1 - F_i E_i C_{\bar{i}} \big], \\[4pt]
		\tau_{i,3} &= F_i \, \tau_{i,1}, \\[4pt]
		\tau_{i,4} &= E_i \, F_i \, \tau_{i,1} + \phi(p_{i,3}) \, \tau_{i,2}.
	\end{align*}
\end{subequations}
with
\begin{subequations}
	\begin{align*}
		A_i &= \phi(p_{\bar{i},2})\big(\phi(p_{i,3}) - 2\big) + B_i,\\
		B_i &= 2\big(1 - \phi(p_{i,3})\big)\,\phi(p_{\bar{i},3}) + \phi(p_{i,3})\,\phi(p_{\bar{i},4}), \\
		C_{\bar{i}} &= \phi(p_{\bar{i},2}) + \phi(p_{\bar{i},4}) - 2\,\phi(p_{\bar{i},3}), \\
		D_i &= \phi(p_{i,5}) - \phi(p_{i,6}), \\
		E_i &= \phi(p_{i,4}) - \phi(p_{i,3}), \\
		F_i &= \frac{\phi(p_{i,6})}{1 - \phi(p_{i,4}) + \phi(p_{i,6})}.
	\end{align*}
\end{subequations}
Substituting \cref{eq_solution_two_body_2} a)-d) into \cref{App_GFs_corrs} g)  allows to
express the residues of the single-particle GF given in \cref{eq:Gr-sixpole} in terms of the local occupations
\begin{subequations}
	\begin{align}
		r_{i,1} =& 1-\braket{n_{i\sigma}}+\left(\tau_{i,1}+2\tau_{i,2}-\tau_{i,3}-2\right)\braket{n_{\bar i\sigma}},\\
		r_{i,2} =& \braket{n_{i\sigma}}+\left(\tau_{i,3}-2\tau_{i,2}\right)\braket{n_{\bar i\sigma}},\\
		r_{i,3} =&2\left(\tau_{i,2}-\tau_{i,3}\right)\braket{n_{\bar i\sigma}},\\
		r_{i,4} =&\tau_{i,3} \braket{n_{\bar i\sigma}},\\
		r_{i,5} =&2\left(1-\tau_{i,1}-\tau_{i,2}+\tau_{i,3}\right)\braket{n_{\bar i\sigma}} ,\\
		r_{i,6} =&\left(\tau_{i,1}-\tau_{i,3}\right)\braket{n_{\bar i\sigma}}.
	\end{align}
	\label{eq_numerators_one_body}
\end{subequations}
\acknowledgments
I am grateful to Stefan Kurth and Fabio Taddei for fruitful discussions. I acknowledge the European Union under the Horizon Europe research and innovation programme (Marie Skłodowska-Curie grant agreement no. 101148213, EATTS)
\bibliography{biblio}

\end{document}